# Cascaded uncoupled dual-ring modulator

**Tingyi Gu[1,2], Young-kai Chen[1], Chee Wei Wong[2] and Po Dong[1]***

[1]*Bell Labs, Alcatel-Lucent, 791 Holmdel Road, Holmdel, NJ 07733 USA*

[2]*Optical Nanostructures Laboratory, Columbia University, New York, NY 10027*

*\*Corresponding author: po.dong@alcatel-lucent.com*

**Abstract:** We demonstrate that by coherent driving two uncoupled rings in same direction, the effective photon circulating time in the dual ring modulator is reduced, with increased modulation quality. The inter-ring detuning dependent photon dynamics, *Q*-factor, extinction ratio and optical modulation amplitude of two cascaded silicon ring resonators are studied and compared with that of a single ring modulator. Experimentally measured eye diagrams, together with coupled mode theory simulations, demonstrate the enhancement of dual ring configuration at 20 Gbps with a *Q* ~ 20,000.

**Introduction:** Silicon photonics modulators have emerged as low-power optical transmitter functions, in combination with energy-efficient CMOS driver amplifiers using hybrid electronic-photonics integration or monolithic integration [1-5]. The silicon modulators were also integrated into an intrachip wavelength division multiplexing (WDM) optical link to form a reconfigurable switch fabric for chip-level optical interconnects and network [6]. Because of their extremely compactness and low energy consumption, silicon microring/microdisks are identified as one of the most energy efficient photonic modulators. High-speed modulation of silicon microring could be obtained by modulating the reverse-biased *pn* junctions embedded in its waveguide. Compared to electrical delay, the photon life time usually sets the upper limit of the cavity modulation rate [4, 7-9]. To reduce the cavity lifetime for higher modulating speed, several schemes were investigated with complex design and layouts. For example, the modulation speed was improved by two times by utilizing the coupling-modulated microrings [10-11]. However, these approaches result in expanded device footprint and require non-conventional fabrication sequences. Dual resonator configuration, such as two rings with a feedback loop or coupled photonic crystal cavities, have been also used to engineer the effective photon lifetime [12-14], in combination with thermal-optic detuning controls [15-16]. In this letter, we show that with a conventional fabrication process and compact device layout, the modulator bandwidth can be improved by introducing cascaded microrings without any feedback. By directing input photons into two microring cavities in series, the resultant effective photon lifetime in the two-ring system can be reduced by half, therefore improving the modulation bandwidth.

Photon dynamics in cascaded uncoupled rings can be described by time domain coupled mode theory (CMT). The photon amplitudes in the first and second resonators are:

$$\frac{da_1}{dt} = \left(i(\omega_1 - \omega) - \frac{1}{2\tau_1}\right)a_1 + \sqrt{\frac{1}{2\tau_{1c}}}s_{1+} \quad (1\text{-}1)$$

$$\frac{da_2}{dt} = \left(i(\omega_2 - \omega) - \frac{1}{2\tau_2}\right)a_2 + \frac{1}{2\tau_{2c}}e^{i(-\beta L)}a_1 + \sqrt{\frac{1}{2\tau_{2c}}}e^{i(-\beta L)}s_{1+} \quad (1\text{-}2)$$

where $a_1$ and $a_2$ are light amplitudes in the first and the second rings. The total photon lifetime ($\tau_{1(2)}$) in a single resonator is $1/\tau_{1(2)}=1/\tau_{1(2)c}+1/\tau_{1(2)in}$, where $1/\tau_c$ and $1/\tau_{in}$ are coupling and intrinsic loss rates respectively. $\omega_{01}$ and $\omega_{02}$ represent the cavity resonant frequencies which are dynamically tunable by carrier depletion to be time dependent as $\omega_1 = \omega_{01} + \Delta\omega$ and $\omega_2 = \omega_{02} + \Delta\omega$, respectively. $\omega$ is the frequency of input laser, and $\Delta\omega$ is the resonance shifted by carrier depletion modulation. $\beta$ is the propagation constant in waveguide and $L$ is the length of waveguide between two rings. $s_{1+}=sqrt(P_{in})$ is the external optical power supply term. The second terms in Eq. 1-1 and third term in Eq. 1-2 represent its coupling to the first and second resonator respectively. The second term in Eq. 1-2 describes photon coupling from the first resonator to the second one with phase delay in waveguide ($\beta L$).

Fig. 1a and b exhibit the transmission spectra of a single (blue) and a dual ring (red) modulator. Two ring resonators are coupled to the same waveguide without a feedback loop (Inset of Fig. 1a). Here, we set single ring resonators with 20,000/80,000 coupling/intrinsic quality factor, and minimal transmission of -14dB. Intrinsic quality factor ($Q_{in}$) is determined by the propagation loss of the waveguide, while coupling quality factor ($Q_c$) depends on the ring-waveguide coupling gap. Both of the quality factors are pre-determined by materials and nanofabrication. The total quality factor ($1/Q_t = 1/Q_{in}+1/Q_c$) is proportional to the photon lifetime ($\tau=Q_t/\omega$). The ratio between $Q_{in}$ and $Q_c$ sets the upper limit of the extinction ratio for the single ring [17]. For a single ring modulator, $Q_c$ can be varied in the design to compromise between extinction ratio and photon lifetime. With the given $Q$s, the extinction ratio (ER) of a single ring modulator is about 7.9dB as the electrical signal drive the resonance shift about half of the full wave half maximum (FWHM). ER is greatly enhanced by dual ring modulator to 20.1dB with zero inter-ring detuning, at the same laser-resonance detuning as a single ring modulator. However, the enhanced insertion loss brings extra power penalty, and may degrade optical modulation amplitude (OMA). OMA can be separately optimized by dual ring configuration, with inter-ring detuning set as 60pm (Fig. 1b). The OMA and ER of a dual ring modulator can be separately optimized at different inter-ring detunings (Fig. 1c).

Quality factors decide the photon lifetime circulating in the ring. As a pulse excitation (delta function) is sent into the waveguide at 0ps, the transmission decay dynamics for single (blue curves) and double-ring configuration (red solid curve) are given in Fig. 1d. Light circulates in a ring for the photon lifetime of 28 ps, and drops to ~11ps when the two rings are coherently driven (Fig. 1d). The effective photon lifetime of a dual ring modulator is insensitive to inter-ring detunings. If the resonance of the two rings are not well aligned (by thermal-optic tuning), a long tail would degrade the extinction ratio and thus the output eyes, while the photon lifetime maintains same as the zero detuning case (dashed curve in Fig. 1d, $\delta=(\lambda_{ring2}-\lambda_{ring2})/(\Delta\lambda/2)=1$). It shows the response time for dual ring modulator has high tolerance for resonance offset between two rings. Slightly different intrinsic quality factors between the two rings lead to the ripples on the tail, and can be eliminated when the quality factors for two rings are identical.

The reduced effective coherently modulated photon lifetime would lead to better high speed modulation quality. The initial laser-resonance detuning is set as 30pm. Given the quality factors for two rings, combined with random pattern input setting, CMT explicit the ideal eye diagram for single ring (Fig. 2a), double ring (Fig. 2b) with push-push scheme ($\delta_{Mod1}= \Delta\lambda/(\Delta\lambda_{Bandwidth1}/2) = 1.5$, $\delta_{Mod2}= 1.5$, Fig. 4b) at 40 Gb/s modulation speed. Dual-ring gives the better eye opening (numerically expressed as $Q$ factor$= (P_1 - P_0)/(\sigma_1 + \sigma_0)$). From the histogram on the eye width, the power level ($P_{1/0}$) and root mean squares ($\sigma_{1/0}$) for one and zero levels can be fit by Gaussian. Here, the simulation does not include the electrical resistance-capacitance times, which in practice would limit the modulation speed as well.

The 'eyes opening' depredates as the clock frequency increases. Dual ring modulator excels in terms of $Q$ factor and ER over single ring for simulated clock frequency ranges from 15GHz to 45GHz, at zero inter-ring detunings. The increased insertion loss slightly lowers the OMA for dual ring modulator (Fig. 2d). The inter-ring detuning is also optimized at 20GHz and 40GHz clock frequencies as shown in Fig. 2e.

To verify the speed enhancement, we use a silicon photonic integrated circuit include 10 ring modulators coupled to a bus waveguide, which was originally used in [6]. The chip was packaged with two fibers with a fiber-to-fiber loss of 13 dB (Figs. 3c). In the current study, we randomly pick two adjacent rings out of the ten. The ring resonance is slowly tuned close to laser resonance and fixed by integrated heaters. The data drive is simultaneously applied on both rings through a dual RF probes (Fig. 3a). The resonance can be tuned by microheater on top of the ring, and the *pn* diodes crossing each ring can fast modulate the ring resonance (Fig. 3b). By placing the heater right on top of the waveguide, the Ohm heat diffuses through the oxide layer to the silicon waveguide, and effectively adjust the laser-cavity detuning.

The steady state characterizations of the rings (transmission and delay) are preceded by Luna Technologies' optical vector analyzer (Fig. 4a) [18], and the measurements are fitted by CMT. The red circles are experimental data and black lines are theoretical fitting. By fitting the transmission and delay spectrum, the coupling/intrinsic quality factors for two rings are 20,000/85,000 and 22,000/100,000, respectively (Fig. 4a). The thermal tuning and electro-optic modulation (reverse bias) are measured to be 80.5 pm/mW and 22pm/V respectively, as summarized from 10 ring modulators with integrated heater and *pn* junctions in [6].

The modulation quality (eye opening) is electrically limited by two factors: Electro-optic modulation depth and the electrical resistance-capacitance constant (electrical delay). As the electrical delay is given by the *pn* junction fabrication, we focus on optimizing the reverse bias dependent transmission modulation and correspondent laser-resonance detunings, where the peak-to-peak driving voltage ($V_{pp}$) is quadratic related to energy consumption per bit [19]. Fig. 4b measures the transmission spectrum of the two rings under reverse bias from 0v to -3v, with 1v per step. The transmission dip (resonance) is effectively shifted from zero detuning ($\delta = 0$) to $\delta = -1.4$ (Lower-right inset of Fig. 4b). The carrier density dependent linear loss modifies the on-resonance transmission near the critical coupling, as the reverse bias red-shifts the resonance [20-21]. The minor modification of ER would not affect our conclusions. The tuning is most efficient as the reverse bias increases from 0V to -2V, in which region we chose as modulation range. The tuning efficiency saturates at higher bias range. A simple quadratic fit for the measured resonance versus reverse bias plot gives ($\lambda_{laser} - \lambda_{res}$) = $-4.5V^2 + 31V + 0.13$ pm. Given the ring radius of 15μm, the RC constant for the single ring modulator is estimated to be ~8ps. For dual ring modulator, the parallel connected pn junctions with half resistance and double capacitance keep the same RC constant as single ring.

The high-speed measurement was carried out by using a dual-head RF probes and two DC probes (Fig. 3c). The resonances of two neighboring rings are independently tuned to be close to laser wavelength at 1563 nm by the integrated heaters, with 30 pm detuning between the ring resonance and laser wavelength. 0-2 V reverse bias modulates ring resonance shifting another 30 pm away from laser wavelength. The high speed performance is measured by optical eye diagrams at 20Gb/s rate. The high speed (20 Gb/s) operations for the first and second rings are recorded on sampling oscilloscope (Fig. 5a and Fig. 5c, with 20 ps per grid). The extinction ratios for the eye diagrams are 5.9 dB and 5.8 dB for single ring operation, and are enhanced to 7.13 dB with double ring co-operation (Fig. 5e), with optimized thermo-optic detunings. Correspondent simulations by CMT with the parameters measured in steady state are presented in Fig. 5b, 5d, 5f for comparison. The eye opening was enhanced as well, matching the theoretical prediction that the fall/rise times are shortened due to reduced effective photon lifetime by coherent modulation.

We have demonstrated that a cascaded dual-ring modulator without feedback can enhance the high speed modulation quality for micro-ring modulators. The simple decay dynamics predicts enhanced photon loss rate when neighboring two identical rings are coherently modulated by the same electrical signal. Experimental demonstration of better eye opening at 20 Gbps operation for double ring is performed by adjusted detuning on the reconfigurable silicon photonic network-on-chip platform, which agrees theoretically by the CMT calculations.

**Acknowledgment:** Part of this project is funded by Intelligence Advanced Research Projects Agency (IARPA) under the SPAWAR contract number N66001-12-C-2011. The views and conclusions contained herein are those of the authors and should not be interpreted as necessarily representing the official policies or endorsements, either expressed or implied, of ODNI, IARPA, or the U.S. Government. The U.S. Government is authorized to reproduce and distribute reprints for Governmental purposes notwithstanding any copyright annotation thereon. We acknowledge support of Dr. Dennis Polla at IARPA, Drs. T.-Y. Liow and Patrick G.-Q. Lo of the Institute of Microelectronics, Singapore on fabrication. T. Gu acknowledges discussions with X.-L. Zhu and C. Chen from Prof. K. Bergman's group at Columbia University, and W. M. J. Green from IBM. T. Gu and C. W. Wong acknowledges partial support from NSF (1069240).

**References**

1. P. Dong, W. Qian, H. Liang, R. Shafiiha, D. Feng, G. Li, J. E. Cunningham, A. V. Krishnamoorthy, and M. Asghari, Opt. Exp. **18**(19), 20298 (2010).

2. X. Zheng, D. Patil, J. Lexau, F. Liu, G. Li, H. Thacker, Y. Luo, I. Shubin, J. Li, J. Yao, P. Dong, D. Feng, M. Asghari, T. Pinguet, A. Mekis, P. Amberg, M. Dayringer, J. Gainsley, H. F. Moghadam, E. Alon, K. Raj, R. Ho, J. E. Cunningham, and A. V. Krishnamoorthy, Opt. Exp. **19** (16), 5172 (2011).

3. M. R. Watts, W. A. Zortman, D. C. Trotter, R. W. Young, and A. L. Lentine, Opt. Exp. **19** (22), 21989-22003 (2011)

4. A. Liu, L. Liao, D. Rubin, H. Nguyen, B. Ciftcioglu, Y. Chetrit, N. Lzhaky, and M. Paniccia, Opt. Exp. **15** (2), 660 (2007)

5. V. M. Passaro, F. Dell'Olio, IEEE Trans. Nanotechnol, **7**(4), 401-408 (2008)

6. P. Dong, Y. K. Chen, T. Gu, L. L. Buhl, D. T. Neilson, J.H. Sinsky, Optical Fiber Communication Conference Th4G.2 (2014).

7. A.C. Turner-Foster, M.A. Foster, J.S. Levy, C.B. Poitras, R. Salem, A.L. Gaeta, and M. Lipson, Opt. Exp. **18**(4), (2010)

8. N.-N. Feng, S. Liao, D. Feng, P. Dong, D. Zheng, H. Liang, R. Shaffiha, G. Li, J. E. Cunningham, A. V. Krishnamoorthy, and M. Asghari, Opt. Exp. **18**(8), 7994 (2010)

9. F. Y. Gardes, A. Brimont, P. Sanchis, G. Rasigade, D. Marris-Morini, L. O'Faolain, F. Dong, J. M. Fedeil, P. Dumon, L. Vivien, T. F. Krauss, G.T. Reed, and J. Marti, Opt. Exp. **17**(24), 21986 (2009)

10. W. D. Sacher, W. M. J. Green, S. Assefa, T. Barwicz, H. Pan, S. M. Shank, Y. A. Vlasov, and J. K. S. Poon, Opt. Exp. 21, 9722 (2013).

11. T. Gu, C. W. Wong, Y. Chen, and P. Dong, In CLEO: Science and Innovations SM2G-1 (2014)

12. Q. Xu, P. Dong and M. Lipson, Nat. Phys. **3**, 406 (2007)

13. S. Kocaman, X. Yang, J. F. McMillan, M. B. Yu, D. L. Kwong, and C. W. Wong, Appl. Phys. Lett. **96**, 221111 (2010)

14. Y. Hu, X. Xiao, H. Xu, X. Li, K. Xiong, Z. Li, T. Chu, Y. Yu, and J. Yu, Opt. Exp. **20**, 15079 (2012)

15. C. Qiu, J. Shu, Z. Li, X.Zhang and Q. Xu, Opt. Exp. **19**(6), 5143 (2011)

16. T. Gu, S. Kocaman, X. Yang, J. F. McMillan, M.-B. Yu, G.-Q. Lo, D.-L. Kwong, C. W. Wong, Appl. Phys. Lett. **98** (12), 121103 (2011)

17. Y. Akahane, T. Asano, B.-S. Song, and S. Noda, Opt. Exp. **13**(4), 1202-1214 (2005)

18. D. K. Gifford, B. J. Soller, M.S.Wolfe, andM. E.Froggatt, Appl. Opt. **44** (34), 7282 (2005)

19. P. Dong, S. Liao, D. Feng, H. Liang, D. Zheng, R. Shafiiha, C.-C. Kung, W. Qiang, G. Li, X. Zheng, A. V. Krishnamoothy, M. Asghari, Opt. Exp. **17** (25), 2248 (2009)

20. A. Yariv. Photon. Tech. Lett., IEEE **14**(4), 483-485(2002)

21. H. Y. Wen, O. Kuzucu, M. Fridman, A. L. Gaeta, L.-W. Luo, and M. Lipson, Phys. Rev. Lett. **108** (22), 223907 (2012)

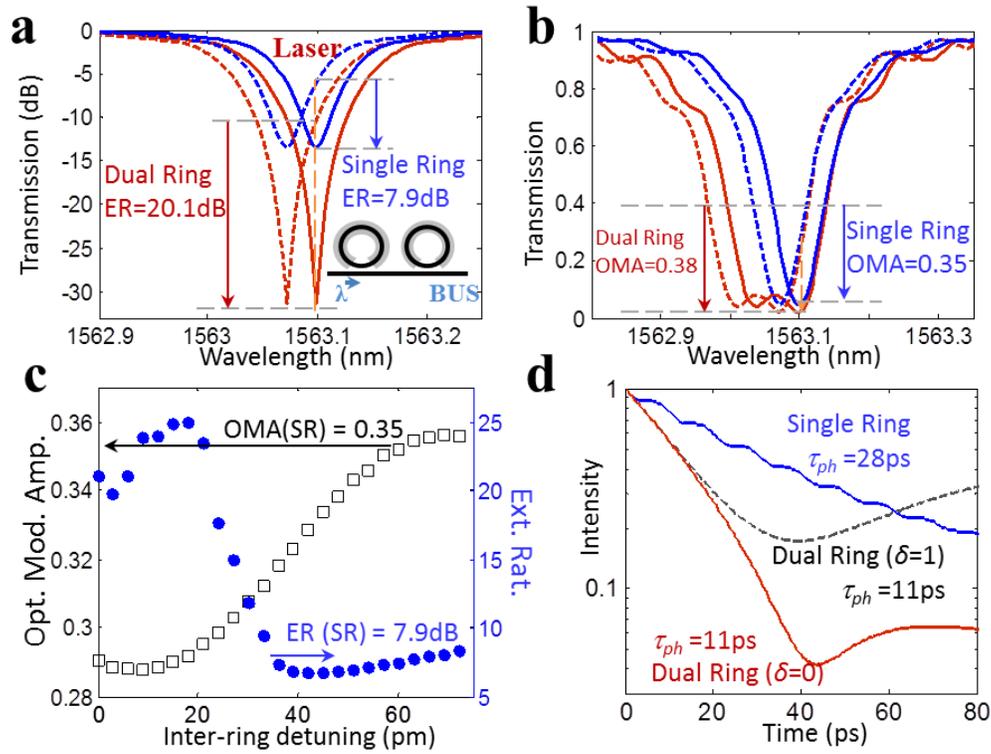

**Fig. 1 Spectral and time domain schematics of a dual ring modulator. a,** Transmission spectrum of typical single (blue) and dual ring (red) modulator, with typical Q(s) of 20,000. The initial state at zero reverse bias (dashed curves) are shifted ~30pm to the right (solid lines). Inset: schematics of dual ring modulator coupled to the same bus waveguide. **b,** Transmission spectrum of single and dual ring modulator with 60pm inter-ring detuning. **c,** OMA and ER versus inter-ring detuning for a dual ring modulator, where the OMA and ER of single ring are marked by arrows. **d,** Photon decay dynamics for single and dual ring.

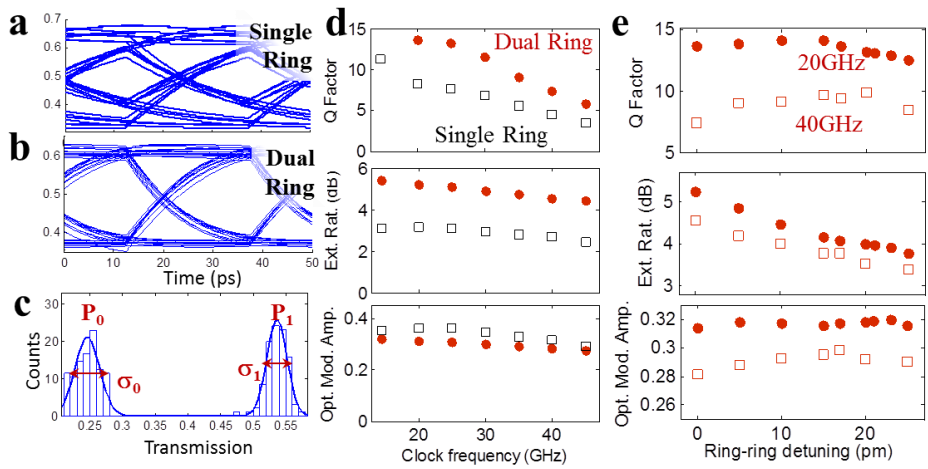

**Fig. 2 High speed modulation performance of a dual ring modulator. a,** Simulated modulation eye diagram for single ring **b,** Double ring operation at 40Gb/s. **c,** Histogram of the transmission along the eye-height. **d,** Q factor, ER and OMA of a dual ring (red solid circle) and single ring (black empty square) modulator with zero inter-ring detuning. **e,** Inter-ring detuning for a dual ring operating at 20GHz (solid circles) and 40GHz (empty squares).

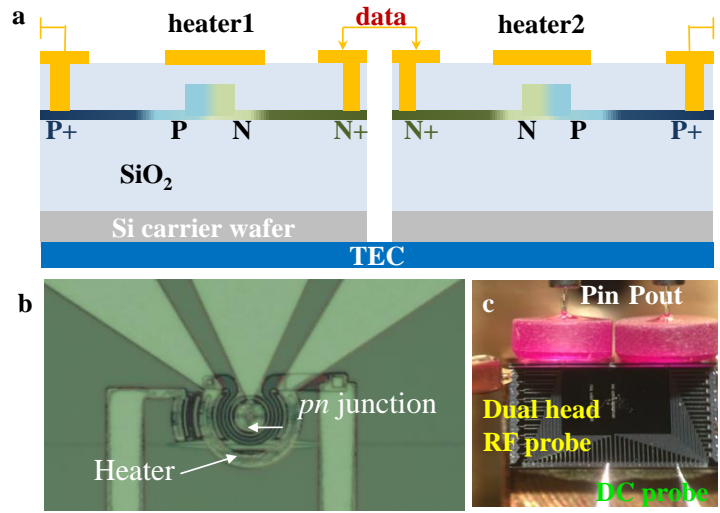

**Fig. 3 Device layout. a,** Schematic cross section drawing of dual ring modulator with lateral *pn* junction and integrated heaters. **b,** Optical top view of the silicon ring with integrated *pn* junction and heater. **c,** Device under test is reconfigurable photonic networks-on-chip based microrings.

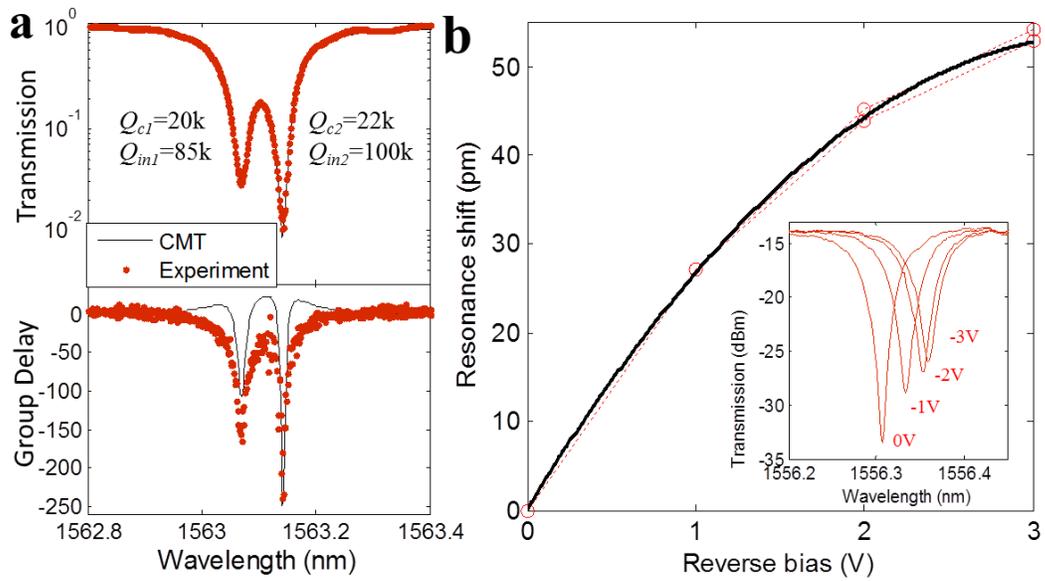

**Fig. 4. DC characteristics. a,** Transmission and delay measurement (red dots) and CMT simulation of double rings. **b,** Resonance shift versus the reverse biased voltage. The dots are experimental data and the black line is quadratic fit. Inset: Transmitted power versus normalized laser-resonance detuning of the ring under reverse bias from 0v to -3v.

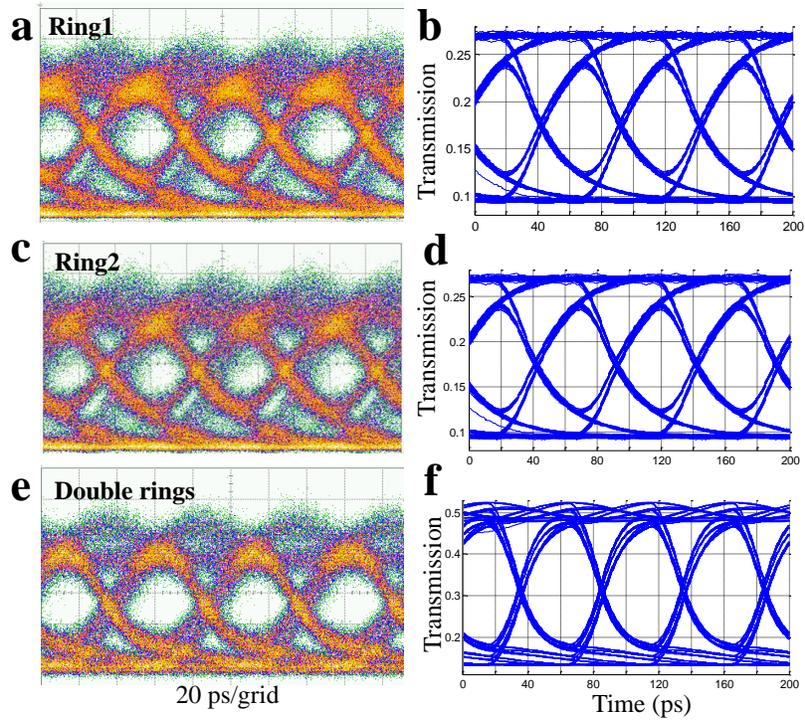

**Fig. 5 Comparison of 20 Gb/s operation for single and dual ring modulators. a,** Measured eye diagrams on the output port of ring 1 on sampling oscilloscope. **b,** Time domain CMT simulation of superposition of 200 single shots, with pseudorandom input. **c,** Experimental **d,** correspondent CMT calculation for ring 2. **e,** Experimental **f,** CMT calculation for double rings with synchronized electrical data drive (considering 10ps electrical delay).